\documentclass{aa}
\usepackage[varg]{txfonts}
\usepackage{graphicx}
\usepackage{natbib}
\bibpunct{(}{)}{;}{a}{}{,}

\begin{document}

\title{Galactic center gamma-ray production by cosmic rays \\ from stellar winds and Sgr A East}

\author{Andrés Scherer\inst{1}
\and Jorge Cuadra\inst{2}
\and Franz E. Bauer\inst{1,3,4}}

\institute{Instituto de Astrofísica, Pontificia Universidad Católica de Chile, Av. Vicuña Mackenna 4860, 7820436 Macul, Santiago, Chile
\and Departamento de Ciencias, Facultad de Artes Liberales, Universidad Adolfo Ibáñez, Av. Padre Hurtado 750, Viña del Mar, Chile
\and Millenium Institute of Astrophysics, Vicuña Mackenna 4860, 7820436 Macul, Santiago, Chile
\and Space Science Institute, 4750 Walnut Street, Suite 2015, Boulder, CO 80301, USA}

\date{Received date / Accepted date }

\abstract
{The High Energy Stereoscopic System (HESS), the Major Atmospheric Gamma-ray Imaging Cherenkov Telescope (MAGIC), and the Very Energetic Radiation Imaging Telescope Array System (VERITAS) have observed diffuse gamma-ray emission strongly correlated with the central molecular zone in the Galactic center. The most accepted scenario to generate this emission is via a hadronic interaction between cosmic rays (CRs) and ambient gas, where CRs are accelerated from a central and continuous source of 1 PeV protons (PeVatron).}{We explore the influence of the three-dimensional (3D) shape of the central molecular zone on the indirect observation of the CR energy density via gamma-ray detection.}{We simulated synthetic gamma-ray maps using a CR diffusion model with spherical injection, one isotropic diffusion coefficient, no advection, and mono-energetic particles of 1 PeV. Also, we used two different 3D gas distributions considering the observed gas column density, both with and without an inner cavity.}{We find that when using a persistent CR source, a disk-like gas distribution is needed to reproduce the existing CR indirect observations. This is in agreement with the continuous gas distribution implied by some dynamical models and studies based on the comparison of emission and absorption molecular lines. However, it contradicts several models of the central molecular zone, which imply that this structure has a significant inner cavity. This tension can be reconciled by an additional, impulsive CR injection.}{If the central molecular zone has a cavity, a composite CR population, coming from the stellar winds of the Wolf-Rayet stars in the central 0.5 pc and the supernova Sgr A East, produces a good match to the observed gamma-ray morphology in the Galactic center.}

\maketitle

\titlerunning{Galactic Center cosmic rays from stellar winds and Sgr A East}
\authorrunning{A. Scherer et. al.}

\section{Introduction}

Very high energy gamma rays (0.1--100 TeV) are a powerful probe of the cosmic-ray (CR) energy density and spectrum.  High energy CR protons collide with low energy protons in the ambient gas, creating neutral $\pi$ mesons, which decay into observable gamma rays. 
The gamma-ray spectrum at high energies adopts a very similar spectrum to its parent proton population, such that gamma rays carry direct information of the high energy proton distribution \citep{Aharonian.2004,Kafexhiu.2014}. 
The gamma-ray luminosity at a given energy is directly proportional to the CR energy density and the gas particle density. 
Therefore, any indirect CR estimates strongly depend on the ambient three-dimensional (3D) gas distribution, a quantity that is usually not easy to obtain.

The Galactic center (GC) is an important source for the production of CRs up to very high energies within the Milky Way \citep{Aharonian.2004b,Aharonian.2009,HESS.2016,HESS.2018,MAGIC.2020,VERITAS.2021}. These CRs interact with the central molecular zone (CMZ), a dense and massive molecular gas region, surrounding the GC on ${\sim}100\,$pc scales, as they diffuse out. In this scenario, a proton interaction produces the diffuse emission strongly correlated with the CMZ morphology, detected by the High Energy Stereoscopic System (HESS), the Major Atmospheric Gamma-ray Imaging Cherenkov Telescope (MAGIC), and the Very Energetic Radiation Imaging Telescope Array System (VERITAS). Also, the inferred CR energy density profile is consistent with a persistent source within the central 10 pc, which is active for over at least $10^4$ years, and the gamma-ray spectrum is coherent with CRs up to 1 PeV (a so-called PeVatron) \citep{HESS.2016,HESS.2018,MAGIC.2020,VERITAS.2021}.
The nature of this source is still unknown. Protons could be accelerating close to the supermassive black hole Sgr A* \citep{HESS.2016}, inside the Arches, Quintuplet, and Nuclear clusters of young massive stars \citep{Aharonian.2019}, or due to an unresolved population of millisecond pulsars \citep{Guepin.2018}. In particular, the stellar winds of the Wolf-Rayet (WR) stars in the central 0.5 pc \citep{Paumard.2006} could continuously accelerate CRs at the shock fronts that formed by their collision \citep{Hinton.2009} and produce 1 PeV protons due to the collective effect of the shock waves created by the entire WR population \citep{Klepach.2000}. In addition to these continuous sources, violent events such as the supernova Sgr A East \citep{Ekers.1983,Maeda.2002} could have generated an impulsive contribution of accelerated particles, in this case at the shock front with its supersonic outflow \citep{Hinton.2009}.

In their analysis, HESS computed the average CR energy density along the lines of sight toward Sgr A* and the Galactic disk indirectly, considering the observed gas column density (i.e., projected on the plane of the sky) in the CMZ and excluding the central $\approx$15 pc [outside the circumnuclear disk (CND), \citealp{Genzel.1988}].  Consequently, their result could be biased by not considering the actual 3D gas density field.  The gas distribution in the GC is difficult to estimate due to the extreme environment of the CMZ: a high temperature gas, a very high molecular density, a clumpy structure, a high level of turbulence, and a strong magnetic field \citep{Battersby.2020}.
Moreover, recent studies have shown that some molecular clouds long assumed to be part of the CMZ are actually in the foreground. For instance, G0.253+0.016 ("the Brick") is argued to be 1 kpc in front of the CMZ \citep{Zoccali.2021}.
The analysis of molecular rotation line emission maps provides the CMZ column density as CO lines \citep{Oka.2012}, CS lines \citep{Tsuboi.1999}, NH$_3$ lines \citep{Purcell.2012,Krieger.2017}, and HCN lines \citep{Jones.2012} act as proxies for the distribution of the dominant yet elusive H$_2$. Using those tracers, several articles have studied the gas 3D distribution and kinematics, for example, searching the orbits that best fit the morphology and radial velocities, modeling the gas features by hydrodynamical simulations, or qualitatively comparing the emission and absorption lines. Recent studies propose different 3D shapes for the CMZ: two spiral arms \citep{Sofue.1995,Ridley.2017}, a twisted elliptical ring \citep{Molinari.2011}, an open elliptical stream \citep{Kruijssen.2015}, either an elliptical ring if the star formation is minimal or a fragmented ring if the star formation is intense \citep{Armillotta.2019,Armillotta.2020}, or a bar-like structure \citep{Sawada.2004,Yan.2017}. The 3D morphology has therefore not been completely determined yet. Moreover, while some of those studies have a continuous or even rising density toward the center \citep{Sawada.2004,Yan.2017,Ridley.2017,Armillotta.2019}, in others there is a large cavity \citep{Sofue.1995,Kruijssen.2015}.

In this work, we analyze the impact of the cavity on the indirect observation of the CR energy density from the CMZ. We create synthetic maps of gamma-ray emission from a CR diffusion model, consistent with the \citet{HESS.2016} constraints, and the CO ($j=3-2$) observations of the CMZ \citep{Oka.2012}. We consider the distribution of the ambient gas along the line of sight in the CMZ to either arise from an elliptical ring or an elliptical disk. 
Finally, we compute the gamma-ray luminosity and the implied CR energy density profiles, so as to compare them with  \citet{HESS.2016} results. We conclude that the data favor a disk shape, in contrast to the studies which suggest an inner cavity. This disagreement can be solved by an additional, impulsive CR source, which we associate to Sgr A East.

\section{Methodology}

We computed a CR diffusion model with continuous and impulsive injection, no advection, and an isotropic diffusion coefficient, using a Monte Carlo method. In all of our models, we included a central continuous source consistent with the WR colliding winds, while in the final one we added a central impulsive source representing Sgr A East. Both sources are located in the central pixel of the simulation, according to their coordinates and our numerical resolution. We tested two different 3D gas distributions, with and without an inner cavity, but both consistent with the observed gas column density. Then we discretized the gas model in a 3D grid of data. The CR diffusion model is independent of the gas density; therefore, the 3D CR distribution was computed separately from the gas distribution. Both ingredients were used to calculate the gamma-ray luminosity in each grid bin and we obtain ed the total gamma-ray surface flux on our line of sight for both gas distributions. Finally, we estimated the ``observable'' average CR energy density by an indirect method, considering the computed synthetic gamma-ray maps and the observed gas column density.

\subsection{Cosmic-ray model} \label{subsec_CR_model}

First, the CR dynamics were computed following the \citet{HESS.2016} analysis. We used a simplified CR diffusion equation for protons, neglecting advection, reacceleration, spallation, and decays \citep{Aharonian.2004,Longair.2011},
\begin{equation}
\label{eq_dif}
\frac{\partial \psi(\vec{r},p,t)}{\partial t}=\vec{\nabla} \cdot (D(E) \cdot \vec{\nabla} \psi)-\frac{\partial}{\partial p}\left(\left(\frac{dp}{dt}\right)\psi\right)+Q(\vec{r},p,t),
\end{equation}
where $\psi(\vec{r},p,t)$ is the CR distribution, $\vec{r}$ is the position vector, $p$ is the total CR momentum at position $\vec{r}$, $t$ is the time, $D(E)$ is the diffusion coefficient, $E$ is the energy of ultrarelativistic particles, and $Q(\vec{r},p,t)$ is the CR source. Equation \ref{eq_dif} was solved by the Monte Carlo method in a 3D domain. In our fiducial models, we considered a constant injection CR source defined by $Q=Q_0E^{-\Gamma}g(t)$ (where $Q_0$ is the CR injection rate, $\Gamma$ is the spectral index of the CR intrinsic spectrum, and $g(t)$ is a Heaviside function in the time interval $0<t$). We simulated mono-energetic particles of 1 PeV from a CR source located at the coordinates of Sgr A*. We simplified the CR intrinsic spectrum because the solution expected by \citet{HESS.2016} from Eq. \ref{eq_dif} was already obtained with mono-energetic particles. In addition, a broader spectrum band would give a similar result when $\psi$ is computed in the relevant energy range (i.e., between 10 TeV and 1 PeV). We set the normalization $Q_0=3.4 \times 10^{36}$ erg s$^{-1}$ to fit the expected CR injection from the WR stellar winds in that region, considering that the total mass-loss rate of the nuclear star cluster WR population is $\sim10^{-3}$ M$_\sun$ yr$^{-1}$, their typical stellar wind velocity is $\approx 1000$ km s$^{-1}$ \citep[see table 1 of][based on \citealt{Martins.2007}]{Cuadra.2008}, and assuming a canonical $1\%$ of the winds' kinetic power going into acceleration of CRs above 10 TeV \citep{Aharonian.2019} (in this simulation, only for 1 PeV protons).

Regarding the CR dynamics, we considered an isotropic $D$ defined as $D(E)\sim 10^{26}(E/10^{10}$ eV$)^\delta$ cm$^2$ s$^{-1}$, and $\delta=0.5$ \citep{Aharonian.2004}, for a Kraichnan spectrum for magnetic turbulence. 
In the GC, this normalization of $D$ is appropriate due to the high turbulent magnetic field associated with the high gas density in this environment \citep{Ormes.1988}, as constrained from gamma-ray observations in other high gas density areas \citep{Abeysekara.2017,Aharonian.2019}. 
It is important to notice that this is two orders of magnitude lower than the standard Galactic interstellar medium (ISM) value used by \cite{HESS.2016}. 
We ran the simulations over a timescale of 10$^6$ years.  This is 100 times longer than the evolution time proposed by the HESS studies \citep{HESS.2016}, which compensates for the fact that we used a diffusion term which is 100 times lower, therefore obtaining the same CR dynamics. Additionally, considering that the CRs are created by the colliding WR winds in the GC, the source is expected to have been roughly constant over that longer timescale \citep{Calderon.2020}, which is approximately the WR population age at the GC \citep{Genzel.2010}. Also, we considered a regime in which the diffusion time is much shorter than the proton-proton energy loss time
(i.e., the first term on the right is much greater than the second term on the right). This assumption is valid for all particle distributions injected, where the average time of the particles within the high-density zone is less than the cooling time in this same area.

We ran ten CR models with 10$^7$ test particles each. The models were discretized on a 3D grid centered in Sgr A*, and we computed the total CR energy inside the volume of each bin ($w_\mathrm{bin}$). The grid covers the range $-1.07\degr<l<1.73\degr$ and $-0.42\degr<b<0.40\degr$ in Galactic longitude and latitude, respectively, and $\pm 132$ pc along the line of sight. For the assumed GC distance of 8.5 kpc, the bins are cubic, with a size of 12 pc, resulting in an angular resolution of $\approx$ 0.08$\degr$.   
The ten simulations were finally averaged over each bin, and we find that the CR profile follows the $w_\mathrm{bin} \propto \psi \propto R^{-1}$ behavior expected for the continuous injection of particles from a central source (see red triangles in Fig. \ref{fig_CR_den_pro}, right panel). 

\subsection{3D gas distribution}

\begin{figure*}
\centering
\includegraphics[width=17cm]{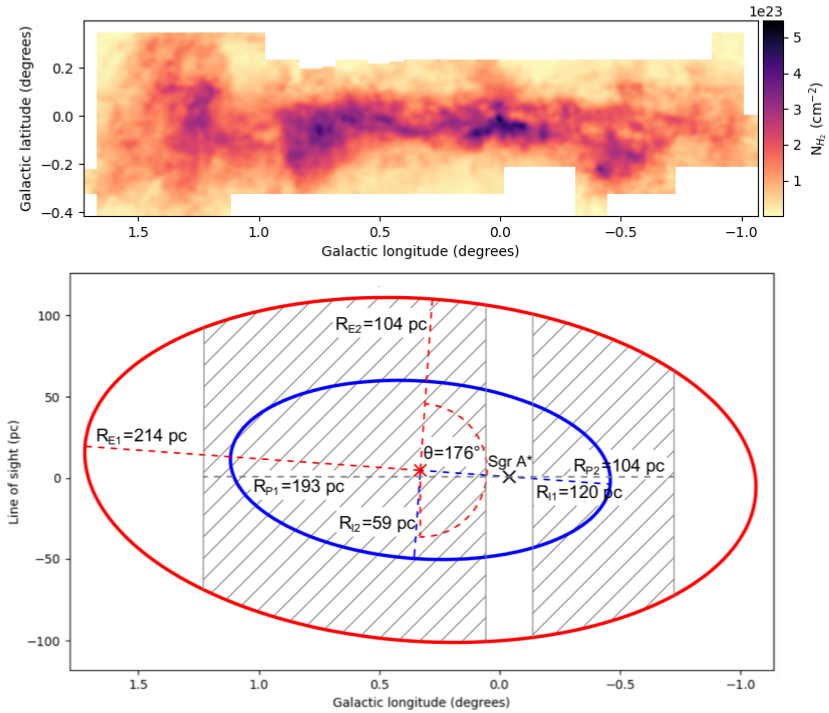}
\caption{CMZ gas distribution. Top panel: CMZ particle column density computed from CO ($J=3-2$). The CMZ was covered between $-1.07\degr<l<1.73\degr$ and $-0.42\degr<b<0.40\degr$, within a velocity range of $|v|< 220$ km s$^{-1}$ \citep{Oka.2012}. The white areas along the perimeter are zones with no observations. Bottom panel: CMZ line-of-sight distribution boundaries for both the disk and ring configurations. The red ellipse is the external boundary for both scenarios, and the blue ellipse is the internal boundary for the ring scenario. We note that R$_\mathrm{E1}$ and R$_\mathrm{E2}$ are the semi-major axis and semi-minor axis for the external ellipse, and R$_\mathrm{I1}$ and R$_\mathrm{I2}$ are the semi-major axis and semi-minor axis for the internal ellipse. Both ellipses are centered on the red x-mark, with an inclination $\theta$ with respect to the line of sight. Gray lines demarcate the area projected along the line of sight analyzed in the HESS analysis, where R$_\mathrm{P1}$ and R$_\mathrm{P2}$ are the maximum projected radii from Sgr A* (black x-mark).}
\label{fig_CMZ_gas}
\end{figure*}

We developed two models to distribute the CMZ gas along the line of sight, considering the CMZ as either an elliptical ring or an elliptical disk. First, the column density of molecular gas of the CMZ was obtained from CO ($J=3-2$) lines observed by the Atacama Submillimeter Telescope Experiment (ASTE) \citep{Kohno.2004}, published by \citet{Oka.2012}.
\footnote{available at \url{https://www.nro.nao.ac.jp/~nro45mrt/html/results/data.html}} The region covered by \citet{Oka.2012} is between $-1.8\degr<l<3.5\degr$ and $-0.8\degr<b<0.9\degr$, with a range in velocity of $|v|< 220$ km s$^{-1}$. The half-power beamwidth of ASTE is 22" for this line. From this region, we selected the area between $-1.07\degr<l<1.73\degr$ and $-0.42\degr<b<0.40\degr$, as mentioned. To compute the particle column density, we used the mass conversion factor CO-to-H$_2$ of $X_\mathrm{CO}=N_\mathrm{H_2}/W_\mathrm{CO}\approx 1.8 \times 10^{20}$ cm$^{-2}$ K$^{-1}$ Km$^{-1}$ s \citep[see table 1 of][]{Bolatto.2013}, with $N_\mathrm{H_2}$ being the H$_2$ particle column density and $W_\mathrm{CO}$ being the integrated brightness temperature of CO ($J=1-0$), as well a ratio of CO($J=3-2$)/CO($J=1-0$) $\approx 0.7$ \citep{Oka.2012} at the GC. The CMZ particle column density obtained is shown in the top panel of Fig. \ref{fig_CMZ_gas} in color scale, as a function of Galactic latitude and longitude.

The bottom panel of Fig. \ref{fig_CMZ_gas} shows the boundaries of both distributions along the line of sight, where the red ellipse is the common external border, and the blue ellipse is the internal border for the ring model. To distribute the gas along the line of sight, we used the ratio between the orbit apocenter and pericenter and the inclination $\theta$ proposed by \citet{Kruijssen.2015} for the CMZ. Using this inclination, the external ellipse semi-major axis (R$_\mathrm{E1}$) was computed from the maximum Galactic longitude extension, and the external ellipse semi-minor axis (R$_\mathrm{E2}$) from the apocenter/pericenter ratio. These distances allowed us to determine the external border for both the disk and ring configurations. For the case of the CMZ modeled as a ring, the internal ellipse was computed considering the torus spatial distributions proposed by \citet{Launhardt.2002} for the CMZ. The torus internal radius proposed by \citet{Launhardt.2002} was considered as the internal ellipse semi-major axis (R$_\mathrm{I1}$), and the internal ellipse semi-minor axis (R$_\mathrm{I2}$) was computed from the apocenter/pericenter ratio. Finally, for each pixel in the column density map, that quantity was uniformly distributed along the line of sight either between the external and internal ellipse for the ring scenario, or inside the external ellipse for the disk scenario.\footnote{We tested nonuniform distributions of the gas and found that the results did not change significantly, see appendix \ref{app-A}.} The ring-disk extension on Galactic latitude is $-0.42\degr<b<0.35\degr$, where zones with no observations are considered to have zero density in all pixels on their line of sight. With this method we obtained two possible realizations of the CMZ 3D morphology, and computed the gas particle density ($n_\mathrm{H}$) in the same data bins of the 3D grid used in \ref{subsec_CR_model}. Given the velocities and timescales involved, we consider this ambient gas to be static.

\subsection{Gamma-ray synthetic maps}

Based on the cosmic energy per volume element, $w_\mathrm{bin}$, computed from the CR model and $n_\mathrm{H}$ obtained from the CMZ 3D distributions, we estimate the gamma-ray luminosity per unit volume ($L_\gamma$/$V_\mathrm{bin}$) in the 3D grid as the following \citep{Fatuzzo.2006,HESS.2016}: 
\begin{equation}
\label{eq_lum}
\frac{L_\gamma}{V_\mathrm{bin}} \approx \kappa_\mathrm{\pi}~\sigma_\mathrm{p-p}~c~\eta_\mathrm{N}~n_\mathrm{H}~w_\mathrm{bin},
\end{equation}
where $\kappa_\mathrm{\pi}$ is the fraction of kinetic energy of high energy protons transferred to $\pi^0$ production, $\sigma_\mathrm{p-p}$ is the cross section for a proton-proton interaction, and $\eta_\mathrm{N}$ is the gamma-ray contribution from heavier nuclei in CRs and ambient gas. For protons with energies in the gigaelectronvolt and teraelectronvolt range, $\kappa_\mathrm{\pi} \approx 0.18$ \citep{Fatuzzo.2006}, for mono-energetic particles of 1 PeV, $\sigma_\mathrm{p-p} \approx 53$ mb \citep{Aharonian.2004}, and $\eta_\mathrm{N} \approx 1.5$ \citep{HESS.2016}. 

\begin{figure*}
\centering
\includegraphics[width=17cm]{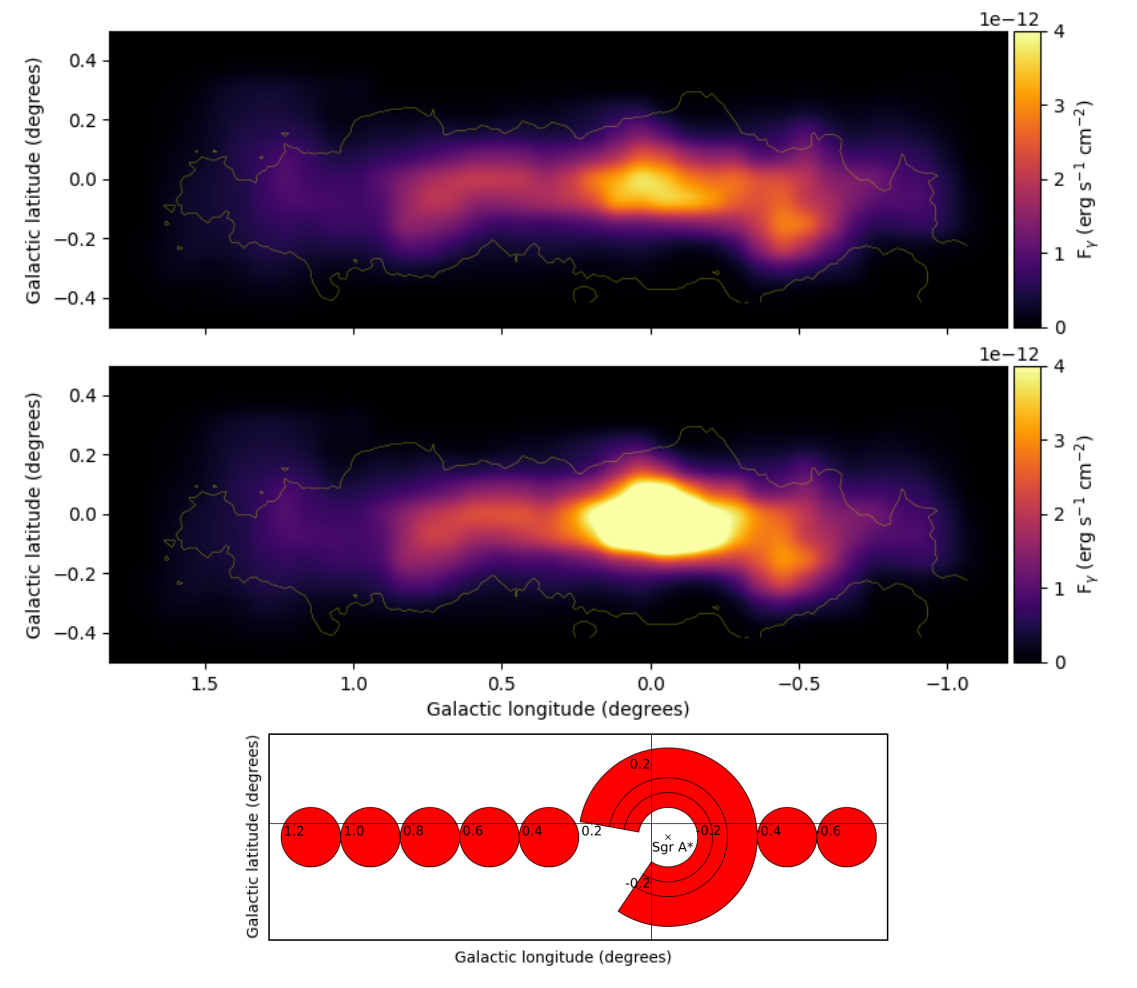}
\caption{Synthetic maps of Gamma-ray flux and analysis areas. Top and middle panels: Gamma-ray flux synthetic maps as observed from Earth for the CMZ ring model (top) and disk model (middle) models. Both maps have been smoothed with a 0.08$\degr$ Gaussian function to adopt the HESS beamwidth. The yellow contour compares the observed gamma-ray area published in the HESS Galactic plane survey \citep{HESS.2018GS}, with the synthetic maps' extension. The contour level is $1 \times 10^{-13}$ erg cm$^{-2}$ s$^{-1}$.
Bottom panel: Regions where we compare our models with the  \citet{HESS.2016} observations in Figs.~3 and 4, shown at the same scale. }       
\label{fig_flux}
\end{figure*}

To obtain the gamma-ray synthetic maps, we integrated $L_\gamma$/$V_\mathrm{bin}$ along each line of sight to calculate the gamma-ray luminosity per bin cross section ($L_\gamma$/$A_\mathrm{bin}$). As we modeled the CMZ as a ring or a disk, we obtained two different gamma-ray maps that are shown in Fig. \ref{fig_flux} as the gamma-ray surface flux computed at the Earth. 
The top panel shows the one corresponding to the CMZ modeled as a ring, while the middle panel shows the map of the CMZ as a disk. Both maps have been adjusted to HESS resolution, that is they have been smoothed with a 0.08$\degr$ Gaussian function to adopt the HESS beamwidth. The grid used is the same as Fig. \ref{fig_CMZ_gas}, but the results are shown in an extension of $-1.20\degr<l<1.85\degr$ and $-0.55\degr<b<0.50\degr$.

\begin{figure*}
\centering
\includegraphics[width=17cm]{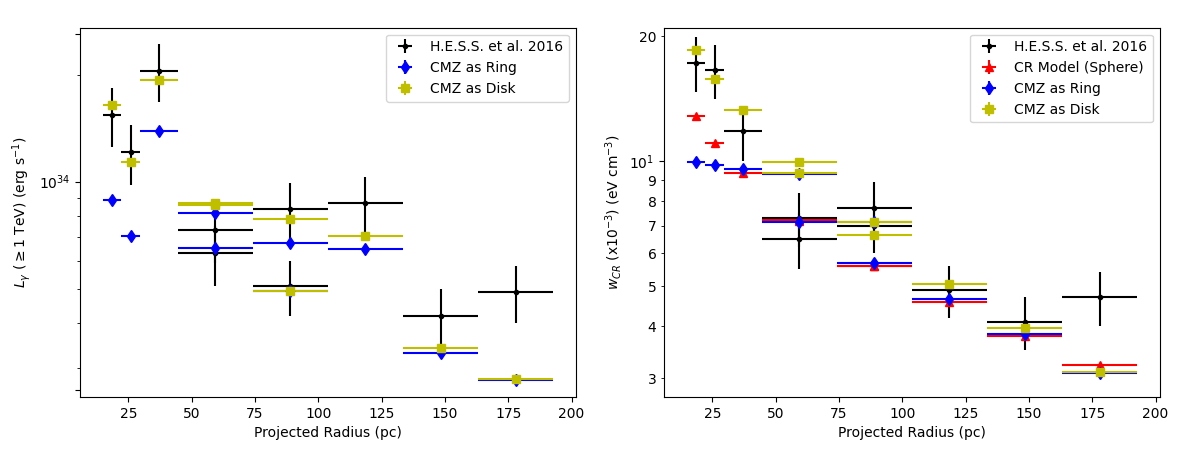}
\caption{Gamma-ray luminosity and average CR energy density from WR stars. Left panel: Gamma-ray luminosity profile associated with the CMZ projected from Sgr A* along the Galactic disk. The profile was binned by integrating the flux in the regions shown in the bottom panel of Fig. \ref{fig_flux}. Right panel: Average CR energy density on the CMZ domain, projected from Sgr A* along the Galactic disk. Black dots denote the observed gamma-ray luminosity (left panel) and the indirect observations of CRs (right panel) computed by \citet{HESS.2016} in the CMZ. Blue diamonds and yellow squares represent the gamma-ray luminosity and the indirect observations of CRs from synthetic maps for the CMZ ring and disk models, respectively. Finally, the red triangles are the intrinsic CR energy density, computed directly from the CR model and considering a spherical domain.}
\label{fig_CR_den_pro}
\end{figure*}

Comparing the maps, both show similar gamma-ray emission far away from Sgr A*, but the disk model has significantly enhanced emission close to the center. 
In order to have a more quantitative comparison between our synthetic maps and HESS observations of this region, we integrated the gamma-ray surface flux over the same regions selected by the HESS study for their profile \citep{HESS.2016}, shown in the bottom panel of Fig. \ref{fig_flux}. 
These correspond to three annular sectors centered in Sgr A* ($l=-0.056\degr$, $b=-0.04588\degr$), with inner and outer radii of 0.1$\degr$ and 0.15$\degr$, 0.15$\degr$ and 0.2$\degr$, 0.2$\degr$ and 0.3$\degr$, and excluding the region between the angles +10$\degr$ and -56$\degr$ from the positive Galactic longitude axis, and seven circular regions with 0.1$\degr$ of radius in $b=-0.04588\degr$ and centered in $l=-0.656\degr$, -0.456$\degr$, 0.344$\degr$, 0.544$\degr$, 359.344$\degr$, 0.744$\degr$, 0.944$\degr$, and 1.144$\degr$.
The comparison of gamma-ray luminosities (Fig.~\ref{fig_CR_den_pro}, left panel) shows that both models reproduce the data between projected radii of $50 \textrm{ pc} \le R \le 150 \textrm{ pc}$ satisfactorily.  
Such a close match is remarkable as no free-parameter normalization was involved in the calculation.
In the region closer to the center, both the data and the CMZ disk model show a large increase, which is not present in the CMZ ring model, suggesting that the CMZ should not have an inner cavity in order to reproduce the gamma-ray observations. At a projected radius of $R\approx175\,$pc, both models predict luminosities a factor of several below the data, which we briefly discuss below. 

\subsection{Synthetic data analysis}

Finally, we compared our results with the CR energy density inferred by HESS. 
For that, we computed the average energy density along the line of sight ($w_\mathrm{CR}$) within the areas shown in the bottom panel of Fig.~\ref{fig_flux}, considering only the gas column density in the CMZ, that is, ignoring the CMZ line-of-sight distribution, as is typically done in the literature. Using Eq. \ref{eq_lum} for the whole CMZ volume, we calculated  $w_\mathrm{CR}$ considering that $n_\mathrm{H}=N_\mathrm{H}/l_\mathrm{los}$ and $V=A~l_\mathrm{los}$, where $A$ is the observed gamma-ray area and $l_\mathrm{los}$ is the length of the domain on the line of sight, as  

\begin{equation}
\label{eq_w_cr}
w_\mathrm{CR} \approx \frac{L_\gamma}{\kappa_\mathrm{\pi}~\sigma_\mathrm{p-p}~c~\eta_\mathrm{N}~N_\mathrm{H}~A}.
\end{equation}

The right panel of Fig. \ref{fig_CR_den_pro} shows the average CR energy density ($w_\mathrm{CR}$) computed from the synthetic gamma-ray maps and compared to the HESS results. Both gamma-ray maps were obtained using the same CR intrinsic energy density profile, which was computed directly from the CR model considering a spherical domain of radius 200 pc, which is roughly the maximum projected radius of these observations (see Fig. \ref{fig_CMZ_gas}). However, when the profiles are inferred indirectly from the observables (in this case, gamma-ray luminosity and gas column density), the intrinsic density profile is not recovered. The synthetic data analysis shows that a cored profile was obtained using the ring model, and a cusped profile was obtained from the disk model.  
Just as was found from the surface brightness comparison, using the inferred $w_\mathrm{CR}$ we can conclude that the CMZ should be a filled disk in order to reproduce the gamma-ray observations.

Finally, the furthest data point presents an energy density increase, which is inconsistent with standard CR transport solutions \citep{Aharonian.2004}. We attribute this discrepancy to more complex CR dynamics at the CMZ border and/or an additional CR source at the location of that specific HESS measurement.

\section{Discussion}

The gamma-ray analysis shows that recovering the intrinsic radial profile of $w_\mathrm{bin} \propto R^{-1}$ for a continuous injection strongly depends on the 3D distribution of the gas surrounding a CR source.
This implies that when a continuous source of CRs is observed, it is not always possible to confirm it robustly or unambiguously by indirect analysis from gamma rays and the gas column density. This issue results from the asymmetric and discontinuous distribution of matter, and it additionally depends on the assumption of CR behavior.

In the specific case of the CMZ, the gamma-ray emission is consistent with the disk model, in contrast to several past dynamical or kinematic studies of gas distribution which argue that the CMZ has a significant inner cavity \citep{Sofue.1995,Kruijssen.2015}, but it is consistent with the continuous distribution inferred from the analysis of emission and absorption lines \citep{Sawada.2004,Yan.2017} and some dynamical studies \citep{Ridley.2017,Armillotta.2019}. If we follow the dynamical studies that propose a cavity, there is a significant inconsistency, which can be solved by modifying our assumptions for the CRs or gamma rays. In particular, below, we consider the possibility of additional gamma-ray sources from the CMZ inner cavity, and also a second impulsive young source of CRs.
An additional centrally peaked gamma-ray luminosity could, in principle, be produced if the region between the CMZ and the CND was characterized by a very low gas density, a weak magnetic field, and a large flux of infrared photons. These photons would collide with high energy electrons that are continuously accelerated in the PeVatron, and generate gamma-ray emission by inverse Compton scattering. For this to work, 100-TeV electrons are needed, that is, an electron energy density similar to protons of 1 PeV, and the far-infrared background should have an energy density of $\approx$ 2 eV cm$^{-3}$. However,  the magnetic field is estimated to be at least 50 $\mu$G \citep{Crocker.2010} in this region, which would generate a fast electron energy loss by synchrotron emission, rendering this inverse Compton scenario infeasible. 

Another alternative to produce the observed gamma rays is dark matter annihilation, expected from a possible cusp in the predicted dark matter density toward the GC \citep{Belikov.2012}. However, the mass and cross section needed to explain the gamma-ray teraelectronvolt emission in the central parsec are too high for most WIMP models of dark matter \citep{Horns.2005}. Therefore, we do not explore this possibility further. 

Our models ignore the Galactic large-scale component of gamma rays on the foreground and background of the CMZ \citep{Abramowski.2014} produced by isotropic Galactic CRs \citep{Blasi.2013}.  However, this large-scale emission is relatively weak and does not contribute in the central region considerably, and hence will not change our main results \citep{HESS.2018}. 

\begin{figure*}
\centering
\includegraphics[width=17cm]{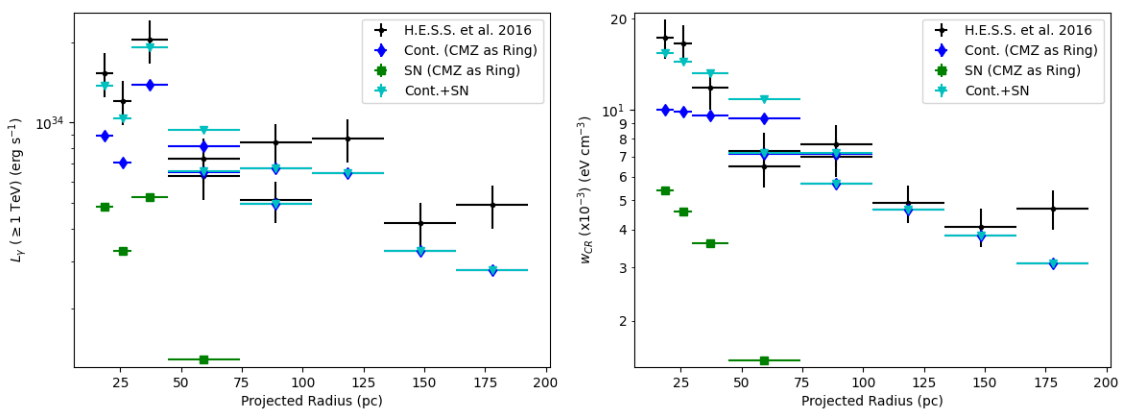}
\caption{ Gamma-ray luminosity and average CR energy density from WR stars and Sgr A East. Left panel: Gamma-ray luminosity from the CMZ in the ring model, projected from Sgr A* along the Galactic disk for a continuous source plus a central impulsive source of CRs, considering the location, injection rate and energy, and age of the WR stars and Sgr A East, respectively. The profile that was binned in the regions is shown in the bottom panel of Fig. \ref{fig_flux}. Right panel: Average CR energy density on the CMZ domain as a ring, projected from Sgr A* along the Galactic disk. Black dots are the observed gamma-ray luminosity (left panel) and the indirect observations of CRs (right panel) computed by \citet{HESS.2016} in the CMZ. Blue diamonds are the gamma-ray luminosity and the indirect observations of CRs from synthetic maps for a CMZ modeled as a ring (same as in Fig.~\ref{fig_CR_den_pro}). Green squares are the gamma-ray luminosity and the indirect observations of CRs from a central impulsive young source for a CMZ modeled as a ring. Cyan triangles are the sum of both components.}
\label{fig_CR_den_pro+SN}
\end{figure*}

We now consider an additional, impulsive young source of protons, which could have created a CR overdensity within the central 50 pc. We refer to the radio source Sgr A East, which is commonly interpreted as a $\sim$1700 yr old supernova remnant, with an explosion energy of $E_\mathrm{SN}\sim1.5\times 10^{51}$ erg \citep{Rockefeller.2005,Fryer.2006}. 
We revised the ring model, using the same configuration described in Sect.~\ref{subsec_CR_model}, that is, based on the continuous WR wind collision, but adding $1\%$ of $E_\mathrm{SN}$ to accelerate protons above 10 TeV \citep{Aharonian.2019}. This impulsive source was simulated as a Dirac delta injection in the time domain and centered on the Sgr A East location, and those particles then diffuse for 1700 yr with the coefficient defined in Sect.~\ref{subsec_CR_model}. Given the slow diffusion, the expected profile of CR energy density around Sgr A East is $w \propto e^{-(R/32.5 \mathrm{~pc}})^2$, rather than the flat distribution that arises for fast diffusion \citep[e.g.,][]{Aharonian.2004}. Figure~\ref{fig_CR_den_pro+SN} shows the gamma-ray luminosity and average CR energy density for the revised model. The sum of both components reproduces the observed gamma-ray emission satisfactorily, without any free parameters. Also the different values for gamma-ray emission at $\approx 60$ pc are noticeable, and they are due to the cavity not being centered on Sgr A*. With this revised model, which includes two physically motivated CR sources, we can reconcile the observations of a gamma-ray central peak with the studies suggesting an inner cavity in the gas distribution.

Finally, if the CMZ is a continuous structure, it is possible to generate the gamma-ray emission only with protons accelerated in the persistent WR wind collision. This constraint is also valid for CMZ distributions with quasi-continuous profiles toward the GC, such as those proposed by \citet{Launhardt.2002} and \citet{Ferriere.2007}. However, in this scenario, the  CR contribution from Sgr A East has to be much lower than our previous assumption; otherwise, the central gas would generate a gamma-ray luminosity peak in the inner 50 pc. 

It does not seem feasible to perform observations to constrain both sources independently at this time.
The angular resolution and sensitivity of the current gamma-ray telescopes are not sufficient to study isolated CR sources in the  central 10 pc, while the Sgr A East variability timescale in low diffusion dynamics can be $\sim$ 100 yr for gamma rays by a proton-proton interaction. 
However, we expect that the GC key science project of the Cherenkov Telescope Array (CTA) \citep{CTA.2019} could confirm or reject any contribution from Sgr A East, and also study the region at $R\approx175\,$pc where a different CR source could be producing enhanced gamma-ray emission.

\section{Conclusion}

We developed a Monte Carlo CR diffusion model for the GC CRs and studied the gamma rays produced by their interaction with the CMZ gas, creating synthetic gamma-ray maps. We considered two different CMZ 3D models, with either a ring or a disk shape. Our simulation results show that if the CMZ has an inner cavity, the gamma-ray luminosity as a function of the projected radius is mostly flat; whereas, if the CMZ has no cavity, a large central gamma-ray spike is produced. The observed gamma-ray emission from the CMZ is consistent with the second option, concordant to emission and absorption line models and some dynamical gas models, but contrary to several kinematic and dynamical gas models that require an inner cavity. 

In order for the gamma-ray morphology to be consistent with a ring CMZ, we propose an additional impulsive cosmic-ray source from the central region. 
Our revised model includes two physically motivated CR sources: a continuous injection of protons accelerated in the stellar wind collisions of WR stars, and an impulsive injection accelerated in the supernova Sgr A East.  The resulting gamma-ray luminosity profile, produced without any free-parameter normalization, matches  the observed data remarkably well.

\begin{acknowledgements} 
We thank the anonymous referee for their useful comments. This project was partially funded by the Max Planck Society through a “Partner Group” grant. AS acknowledges the help and useful comments by Brian Reville, the hospitality of the Max Planck Institute for Nuclear Physics, where part of the work was carried out, and funding from the Deutscher Akademischer Austauschdienst (DAAD). 
JC acknowledges financial support from FONDECYT Regular 1211429.
FEB acknowledges support from ANID-Chile Basal AFB-170002, FONDECYT Regular 1200495 and 1190818, and Millennium Science Initiative Program  – ICN12\_009.
The Geryon cluster at the Centro de Astro-Ingenieria UC was extensively used for the calculations performed in this paper. BASAL CATA PFB-06, the Anillo ACT-86, FONDEQUIP AIC-57, and QUIMAL 130008 provided funding for several improvements to the Geryon cluster.

\end{acknowledgements}

\bibliographystyle{aa}
\bibliography{Bibliography}

\begin{appendix}
\section{Nonuniform distribution tests} \label{app-A}
\begin{figure}[h]
\resizebox{\hsize}{!}{\includegraphics{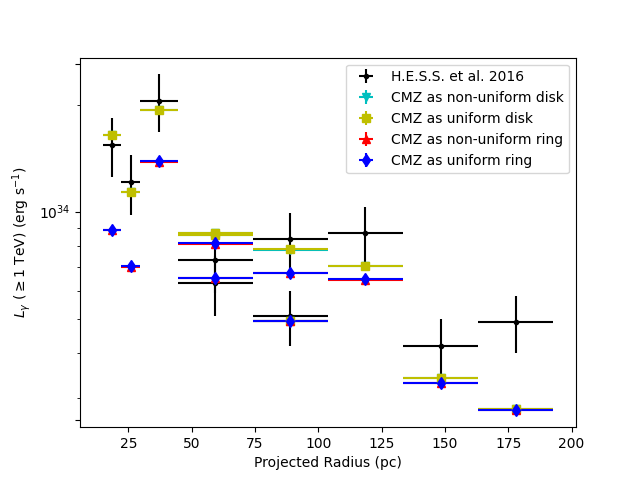}}
\caption{Gamma-ray luminosity from the CMZ projected from Sgr A* along the Galactic disk for a continuous source. The profile was binned in the regions shown in the bottom panel of Fig. \ref{fig_flux}. Black dots denote the observed gamma-ray luminosity. Cyan triangles, yellow squares, red triangles, and blue diamonds are the gamma-ray luminosity from the CMZ as a nonuniform disk, uniform disk, nonuniform ring, and uniform ring, respectively.  Most cyan triangles are behind the corresponding yellow squares, just as the red triangles and blue diamonds.}
\label{fig_app}
\end{figure}

Considering the densities and coordinates of the dust ridges b, c, d, e, and f described in \citet{Walker.2018} and their location along the line of sight proposed by \citet{Kruijssen.2015}, we tested CMZ nonuniform distributions. For the ring and disk shapes, we put the  molecular gas density associated with the dust ridges on the corresponding data bin and distributed the difference between this density and the total observed density uniformly along the line of sight delimited by Fig.~\ref{fig_CMZ_gas}. Figure~\ref{fig_app} compares the data analysis from synthetic gamma-ray maps for a nonuniform and uniform distribution. The difference between the corresponding pairs of models is negligible. In conclusion, the overdensities on the line of sight related to the dust ridges b, c, d, e, and f do not noticeably change the gamma-ray results.

\end{appendix}
\end{document}